\documentclass[Conference,a4paper]{IEEEtran}
\IEEEoverridecommandlockouts
\usepackage{color}
\usepackage{graphicx}
\usepackage{epstopdf}
\usepackage{amsmath}
\usepackage{amssymb}
\usepackage{algorithm}
\usepackage{algorithmic}
\usepackage{amsmath}
\usepackage{multirow}
\usepackage{booktabs}
\usepackage{array}
\usepackage{amsthm}
\usepackage{stfloats}
\usepackage{caption}
\usepackage{subfigure}
\usepackage{bm}
\usepackage{booktabs}
\usepackage{setspace}

\newcommand{\be}{\begin{equation}}
\newcommand{\ee}{\end{equation}}
\newcommand{\bea}{\begin{eqnarray}}
\newcommand{\eea}{\end{eqnarray}}
\newcommand{\ba}{\begin{array}}
\newcommand{\ea}{\end{array}}

\title{Deep Reinforcement Learning based Joint Active and Passive Beamforming Design for RIS-Assisted MISO Systems
\thanks{This work is supported in part by the National Natural Science Foundation of China (Grant No. 61971088, 62071083, 62071105, U1808206, and U1908214), the Natural Science Foundation of Liaoning Province (Grant No. 2020-MS-108), in part by the Fundamental Research Funds for the Central Universities (Grant No. DUT20GJ214, DUT21GJ208 and DUT20RC(3)029), in part by Dalian Science and Technology Innovation Project (Grant No. 2020JJ25CY001),  and  in part by the Open Research Fund of National Mobile Communications Research Laboratory, Southeast University (Grant No. 2021D08).}}
\author{\IEEEauthorblockN{Yuqian Zhu$^{\dag}$, Zhu Bo$^{\dag}$, Ming Li$^{\dag}$$^{\ddag}$, Yang Liu$^{\dag}$, Qian Liu$^{\dag}$, Zheng Chang$^{*}$, and Yulin Hu$^{\sharp}$
\vspace{-0.0 cm} }\\ 
\IEEEauthorblockA{$^{\dag}$ Dalian University of Technology, Dalian, Liaoning 116024, China \\ E-mail: \texttt{\{yqzhu,zhubo\}@mail.dlut.edu.cn, \{mli,yangliu\_613,qianliu\}@dlut.edu.cn} } \\
\IEEEauthorblockA{$^{\ddag}$ National Mobile Communications Research Laboratory Southeast University, Nanjing, Jiangsu 210096, China} \\
\IEEEauthorblockA{$^{*}$ University of Electronic Science and Technology of China, Chengdu, Sichuan 611731, China \\ E-mail: \texttt{zheng.chang@jyu.fi} }\\
\IEEEauthorblockA{$^{\sharp}$ Wuhan University, Wuhan, Hubei 430072, China \\ E-mail: \texttt{yulin.hu@whu.edu.cn} }
}

\begin{document}

\maketitle
\pagestyle{empty}
\thispagestyle{empty}

\begin{abstract}
Owing to the unique advantages of low cost and controllability, reconfigurable intelligent surface (RIS) is a promising candidate to address the blockage issue in millimeter wave (mmWave) communication systems, consequently has captured widespread attention in recent years.
However, the joint active beamforming and passive beamforming design is an arduous task due to the high computational complexity and the dynamic changes of wireless environment. In this paper, we consider a RIS-assisted multi-user multiple-input single-output (MU-MISO) mmWave system and aim to develop a deep reinforcement learning (DRL) based algorithm to jointly design active hybrid beamformer at the base station (BS) side and passive beamformer at the RIS side.
 By employing an advanced soft actor-critic (SAC) algorithm, we propose a maximum entropy based DRL algorithm, which can explore more stochastic policies than deterministic policy, to design active analog precoder and passive beamformer  simultaneously. Then, the digital precoder is determined by minimum mean square error (MMSE) method.
The experimental results demonstrate that our proposed SAC algorithm can achieve better performance compared with conventional optimization algorithm and DRL algorithm.
\end{abstract}


\maketitle

\section{Introduction}
Recently, wireless communication networks need to expand the capacity to meet the exponentially increasing high-data-rate requirements \cite{7446253}. Many newly emerged technologies are employed to increase the capacity of the wireless channels. One of the key enabling techniques is millimeter wave (mmWave) communications associated with massive multiple-input multiple-output (MIMO) and hybrid beamforming techniques. However, the blockage issue makes the mmWave MIMO communications extremely challenging for the real-world deployment.

The reconfigurable intelligent surface (RIS), as an environmentally friendly, low-cost, and controllable planar array, has been considered as one of the vital technologies to tackle the challenge \cite{9326394}, \cite{8910627}. The emergence of RIS has benefited from advancement in the electromagnetic (EM) meta-material, which can control the propagation environment of EM waves in the wireless communication systems. In addition, RIS can establish virtual links to cover signal blind areas and enhance communication quality of cell-edge users.

Many researches have investigated the effective algorithms of active beamforming and passive beamforming design in RIS assisted wireless communication system \cite{8855810}-\cite{xiu2021sumrate}.
In \cite{8855810}, a point-to-point RIS-assisted MISO communication system is investigated. The authors proposed fixed point iteration and manifold optimization methods to maximize the spectral efficiency.
The authors in \cite{9148947} proposed alternating optimization (AO) and semi-definite relaxation (SDR) algorithms to optimize the beamforming vector at the base station (BS) and the phase-shifts at the RIS with imperfect channel state information (CSI).
In \cite{21}, the weighted sum-rate problem was decoupled via Lagrangian dual transform. Then, the transmit beamforming was optimized by the fractional programming
method, and the passive beamforming at RIS was optimized by three efficient algorithms with closed-form expressions.
In \cite{xiu2021sumrate}, the authors employed AO, successive convex optimization (SCA), and SDR algorithms to obtain the optimize solution of active beamforming and discrete phase-shift matrix. However, it is difficult for these aforementioned approaches to accurately estimate the channel in real-world deployment. Also,  the iterative algorithms have inevitably huge computational complexity which introduce unnegligible processing delays.

The artificial intelligence (AI) techniques can efficiently solve massive data, mathematically difficult non-linear and non-convex problems.
Deep reinforcement learning (DRL) as one of the powerful AI techniques has been considered as a promising candidate to handle the dynamic adaption problem in complicated environment.
Compared to the deep learning (DL) approaches, the DRL technique does not require a large amount of training data, which might be very difficult to obtain in wireless communication systems. The DRL based approaches can continuously seek for the optimal combination policy of beamforming design by observing the reward value in time-varying environment without the priori knowledge, e.g., the channel model and the user movement pattern. Thus, the DRL based approach is more capable of handling beamforming design problem in time-varying wireless communication systems \cite{8938771}-\cite{9322372}.
In \cite{8938771}, the authors proposed deep Q-learning (DQN) algorithm with its greedy nature to joint design beamforming, power control, and interference coordination. The authors designed the binary coding to execute the action of  agent, control the BS power and the beamforming codebook. 
In \cite{9110869}, the active beamforming and passive beamforming are jointly designed to maximize the sum rate utilizing deep deterministic policy gradient (DDPG), in contrast to solving the discrete action space. The action space is simply designed by the beamforming matrix and the phase-shift matrix.
In \cite{zhang2020millimeter}, the authors proposed a distributional RL to learn the optimal passive beamforming of RIS in the imperfect CSI scenario.
In \cite{8968350}, the authors introduce DDPG to optimize the passive phase shift at RIS.
Furthermore, the authors in \cite{9322372} formulate a robust power minimization problem considering the RIS's power budget constraint and receiver's signal-to-noise ratio (SNR) requirement. When part of actions are generated by the DDPG algorithm, the rest of actions are obtained by the model-based convex approximation. However, the above algorithms are not effective in optimizing large-scale continuous variables.

Motivated by the above analysis, in this paper, we utilize an off-policy, soft actor-critic (SAC) algorithm to solve the joint beamformer design problem.
Particularly, we consider a RIS-assisted multi-user multiple-input single-output (MU-MISO) mmWave system and aim to design a SAC algorithm based on the maximum entropy  DRL framework to jointly design active hybrid beamformer at the BS and passive beamformer at the RIS.
The proposed SAC algorithm, which can maximize the reward and the entropy by exploring more stochastic policies, jointly designs active analog precoder and passive beamformer. Then, the digital precoder is designed by minimum mean square error (MMSE) method.
The experimental results demonstrate that our proposed SAC algorithm can achieve  better performance compared with conventional optimization algorithm and DRL algorithm.

\section{System Model and Problem Formulation}
\vspace{0.1 cm}
\subsection{System Model}
\begin{figure}[t]
\centering
\includegraphics[width = 3.0 in]{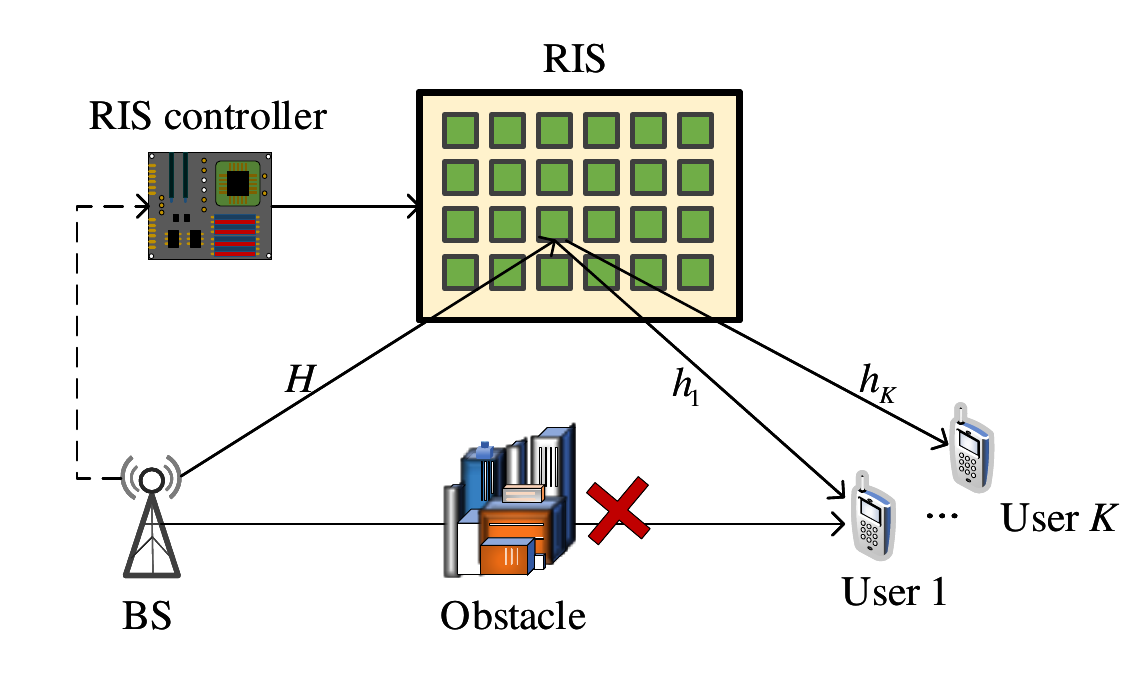}
\vspace{-0.1 cm}
\caption{A RIS-assisted MU-MISO system.}
\label{fig:model}
\vspace{-0.4 cm}
\end{figure}

\begin{figure}[t]
\centering
\includegraphics[width = 2.5 in]{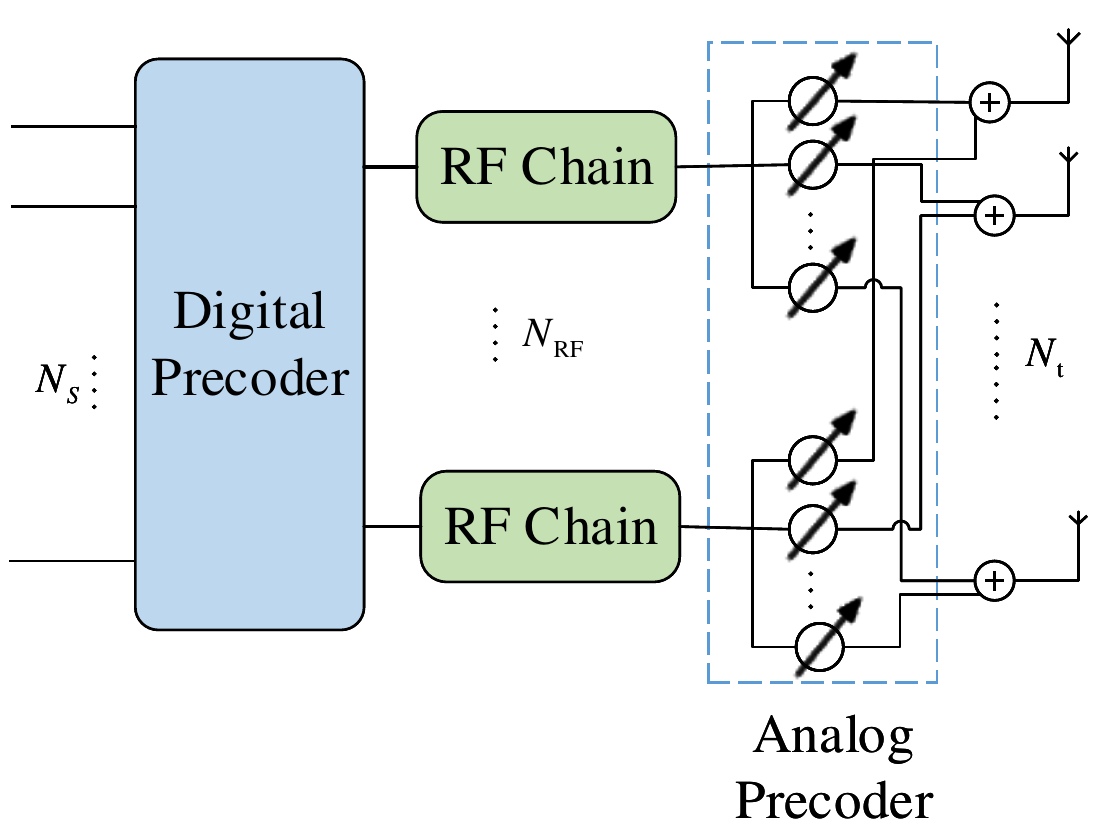}
\vspace{-0.1 cm}
\caption{The specific architecture of hybrid precoder.}
\label{fig:bs}
\vspace{-0.6 cm}
\end{figure}

 As shown in Fig. \ref{fig:model}, we consider a RIS-assisted mmWave multi-user MISO system, where a BS equips with $N_{\mathrm{t}}$ antennas and $N_{\mathrm{RF}}$ RF chains to simultaneously transmit $N_\mathrm{s}$ data streams to serve $K$ single antenna users with the assistance of a RIS of $M$ reflecting elements. To achieve the maximum spectrum efficiency, we assume $K=N_{\mathrm{RF}}=N_\mathrm{s}$.
  The transmitted symbols are first processed by a baseband digital precoder $\mathbf{F}_\mathrm{B B} \triangleq \left[\mathbf{f}_{\mathrm{B B},{1}}, \ldots, \mathbf{f}_{\mathrm{B B},{K}}\right]\in \mathbb{C}^{N_\mathrm{RF} \times K}$, and then up-converted to the RF domain via $N_{\mathrm{RF}}$ RF chains before being precoded with an analog precoder $\mathbf{F}_\mathrm{R F}(i, j) =\frac{1}{\sqrt{N_\mathrm{t}}} e^{j \theta_{i, j}}, \theta_{i, j} \in[0,2 \pi)$ of dimension ${N_\mathrm{t} \times N_\mathrm{R F}}$. The specific details of the hybrid beamforming architecture are shown in Fig. \ref{fig:bs}.

  In addition, we assume that the direct links are blocked by obstacles.
  Denote the phase-shift matrix introduced by the RIS as $\mathbf{\Phi} \triangleq \operatorname{diag}\left\{\phi_{1}, \phi_{2}, \ldots, \phi_{M}\right\} \in \mathbb{C}^{M \times M}$, where $\boldsymbol{\Phi}(m, m)=\phi_{m}=\chi_{m} e^{j \varphi_{m}}$, $m=1,2,\ldots, M$, $\chi_{m}\in[0,1]$ and $\varphi_{m} \in[0,2\pi)$ are the amplitude and phase-shift of each RIS element, respectively. Considering that the RIS is a passive device, we assume that $\chi_{m}=1$. Then, the received signal at the $k$-th user can be written as:
\begin{equation}
\small
y_{k}={\sqrt{P}\mathbf{h}_{k}^{H} \mathbf{\Phi} \mathbf{H} \mathbf{F}_\mathrm{R F} \mathbf{f}_{\mathrm{B B},{k}} s_{k}}
+\sqrt{P}{\sum_{{i} \neq {k}}^{K} \mathbf{h}_{k}^{H} \mathbf{\Phi} \mathbf{H} \mathbf{F}_\mathrm{R F} \mathbf{f}_{\mathrm{B B},{i}} s_{i}}+n_{k},
\end{equation}
where $s_{k}$ is the transmitted symbol for the $k$-th user, $P$ represents transmit power and the power constrains are limited by normalizing $\mathbf{F}_\mathrm{B B}$ such that $\left\|\mathbf{F}_\mathrm{R F} \mathbf{F}_\mathrm{B B}\right\|_{F}^{2}=N_\mathrm{s}$, $n_{k}$ is the additive white Gaussian noise (AWGN) at the $k$-th user with zero mean and noise variance $\sigma^{2}$, i.e. $n_{k} \sim \mathcal{C N}\left(0, \sigma^{2}\right)$.
  In addition, the channels from the BS to the reflecting RIS and from the reflecting RIS to the $k$-th user are denoted by $\mathbf{H} \in \mathbb{C}^{M \times N_\mathrm{t} }$ and $\mathbf{h}_{k} \in \mathbb{C}^{M \times 1}$, respectively.
We adopt the classic geometric channel model,
the channels $\mathbf{H}$ from the BS to the RIS and $\mathbf{h}_{k}$ from the RIS to the $k$-th user can be simply expressed as \cite{channel}
\begin{equation}
\small
\mathbf{H}=\sqrt{\frac{N_{\mathrm{t}} M}{L}} \sum_{l=1}^{L} \alpha_{l} \mathbf{a}_\mathrm{A}\left(N_{\mathrm{t}}, \phi_{l}\right) \mathbf{a}_\mathrm{R}^{T}\left(M, \theta_{a l}, \varphi_{a l}\right),
\end{equation}
\begin{equation}
\small
\mathbf{h}_{k}=\sqrt{\frac{N_{\mathrm{t}}}{L}} \sum_{l=1}^{L} \alpha_{k, l} \mathbf{a}_\mathrm{R}\left(M, \theta_{dl}, \varphi_{dl}\right), \forall k,
\end{equation}
where $L$ denotes the number of multipaths, $\alpha \sim \mathcal{C} \mathcal{N}(0,1)$ is the complex gain, $\mathbf{a}_\mathrm{A}\left(N_{\mathrm{t}}, \phi_{l}\right)\in \mathbb{C}^{N_{\mathrm{t}} \times 1}$ and $\mathbf{a}_\mathrm{R}\left(M, \theta_{a l}, \varphi_{a l}\right)\in \mathbb{C}^{N \times 1}$ represent array steering vectors at BS and RIS, respectively. $\phi_{l}$ is the angle of departure (AoD) of the $l$-th path at the BS, $\theta_{a l}$, $\varphi_{al}$, $\theta_{dl}$ and $\varphi_{dl}$ denote the angles of arrival (AoAs) in horizon and vertical, and the AoDs in horizon and vertical of the $l$-th path at the RIS, respectively. For the $N_\mathrm{t}$-elements array antenna at the BS, the array steering vector can be written as
\begin{equation}
\small
\mathbf{a}\left(N_\mathrm{t}, \phi_{l}\right)= \frac{1}{\sqrt{N_\mathrm{t}}}\left[1, e^{-j \frac{2 \pi}{\lambda} d_{0} \cos \phi_{l}}, \ldots, e^{-j \frac{2 \pi}{\lambda}(N_\mathrm{t}-1) d_{0} \cos \phi_{l}}\right]^{T},
\end{equation}
where $d_{0}$ is the antenna spacing and $\lambda$ is the mmWave wavelength. The array steering vector of the RIS is $\mathbf{a}_\mathrm{R}(M, \theta, \varphi)=\mathbf{a}\left(M_{a z}, \theta\right) \otimes \mathbf{a}\left(M_{e l}, \varphi\right)$.

\subsection{Problem Formulation}
The sum-rate of the RIS assisted MU-MISO system is given by
\begin{equation}
\small
R=\sum_{k=1}^{K} \log _{2}\left(1+\Upsilon_{k}\right),\label{eq:R}
\end{equation}
where $\Upsilon_{k}$ is the signal-to-interference-plus-noise ratio (SINR) of the $k$-th user, which can be expressed as
\begin{equation}
\small
\Upsilon_{k}=\frac{{P}\left|\mathbf{h}_{k}^{H} \mathbf{\Phi} \mathbf{H} \mathbf{F}_\mathrm{R F} \mathbf{f}_{\mathrm{B B},{k}}\right|^{2}}{{P}\left|\sum_{{i} \neq {k}}^{K} \mathbf{h}_{k}^{H} \mathbf{\Phi} \mathbf{H} \mathbf{F}_\mathrm{R F} \mathbf{f}_{\mathrm{B B},{i}}\right|^{2}+{\sigma_{k}}^{2}}, \forall k.
\end{equation}

 We aim to jointly design the optimal digital beamformer $\mathbf{F}_\mathrm{B B}$, analog beamformer $\mathbf{F}_\mathrm{R F}$ and phase-shift matrix $\mathbf{\Phi}$ of RIS that maximize sum-rate of the RIS assisted MU-MISO system. The optimization problem can be formulated as
\begin{equation}
\begin{aligned}
&\max _{\left\{\mathbf{F}_\mathrm{R F}, \mathbf{F}_\mathrm{B B}, \mathbf{\Phi}\right\}} \quad R=\sum_{k=1}^{K} \log _{2}\left(1+\Upsilon_{k}\right) \\
&\quad\quad\text { s.t. } \quad\quad \mathbf{F}_\mathrm{R F}(i, j) =\frac{1}{\sqrt{N_\mathrm{t}}} e^{j \theta_{i, j}}, \theta_{i, j} \in [0,2\pi), \\
&\quad\quad\quad\quad\quad\quad\| \mathbf{F}_\mathrm{R F} \mathbf{F}_\mathrm{B B} \|_{F}^{2}=N_\mathrm{s}, \\
&\quad\quad\quad\quad\quad\quad |\phi_{m}|=1, \forall m=1,2, \ldots, M.
\end{aligned}
\end{equation}
Obviously, the above optimization problem is an NP-hard problem and is difficult to solve by the conventional optimization methods due to the non-convex constraint.
 Since our goal is to design the high-dimensional continuous variables of the analog precoder and phase-shift of RIS, some classic DRL algorithms, such as DQN, DDPG, cannot handle the variables efficiently and often provide a poor local-optimum. Therefore,
  we employ a SAC-based DRL algorithm for the joint active and passive beamforming design.

\section{SAC-Based Joint Hybrid and Passive Beamforming Design}
In this section, we formulate the joint active and passive beamforming design as a markov decision process (MDP) problem and propose a SAC algorithm in the DRL framework to seek the solution of this problem.
Firstly, we introduce the MDP problem formulation,  the mechanism and update  strategy of SAC. After designing active analog beamformer $\mathbf{F}_\mathrm{R F}$ and passive beamformer $\mathbf{\Phi}$ in each learning step,  $\mathbf{F}_\mathrm{B B}$ is obtained by MMSE method. The details are described as follows.

\subsection{MDP Problem Formulation}
We aim to find update policies, which allow the baseband digital beamformer at the BS and the RIS to reasonably update content items under different states by maximizing the long-term average reward. We model the BS and the RIS as an agent. The action of the agent is the variable to be optimized, and the sum-rate is maximized by finding the largest reward. Then, we define the basic elements of the agent MDP as follows.

\begin{itemize}
\item Action $\boldsymbol{a}_{t}$: The action in the RIS-assisted MU-MISO communication system consists of  the phase-shifts at the analog beamformer and the RIS. Thus, the action is expressed as
\begin{equation}
\boldsymbol{a}_{t}=\left[\theta_{1,1}^{(\mathrm{t})}, \ldots, \theta_{K, N_\mathrm{R F}}^{(t)}, \phi_{1}^{(t)}, \ldots, \phi_{M}^{(t)}\right].
\end{equation}

\item State $\boldsymbol{s}_{t}$: The state of the system mainly consists of three parts, i.e., the action at time $t-1$, the channel $\mathbf{H}$ between the BS and the RIS and the channel $ \mathbf{h}_{k}$ between the RIS, and the $k$-th user at time $t-1$. Then, we define the state of the $t$-th step as
\begin{equation}
\boldsymbol{s}_{t}=\left[\boldsymbol{a}^{(t-1)}, \mathbf{H}^{(t-1)}, \mathbf{h}_{1}^{(t-1)}, \ldots, \mathbf{h}_{K}^{(t-1)}\right].
\end{equation}

\item Reward $\boldsymbol{r}_{t}$: The objective is to maximize the achievable rate. Thus, the achievable rate defined in (\ref{eq:R}) is used as the reward function:
    \begin{equation}
    \boldsymbol{r}_{t} = R.
    \end{equation}
\end{itemize}
\vspace{-0.7 cm}
\subsection{Mechanism of Soft Actor-Critic Learning}
We utilize SAC to update policies that maximize the reward in the dynamic environment.
SAC is an advanced off-policy, actor-critic, and entropy-based DRL algorithm \cite{haarnoja2018soft}. Unlike the traditional DRL strategy that only seeks the maximum of the expected sum of rewards, i.e., $\sum_{t} \mathbb{E}_{\left(\boldsymbol{s}_{t}, \boldsymbol{a}_{t}\right) \sim \rho_{\pi}}\left[r\left(\boldsymbol{s}_{t}, \boldsymbol{a}_{t}\right)\right]$, SAC scheme also takes the expected entropy objective to adopt stochastic policies over $\rho_{\pi}\left(\boldsymbol{s}_{t}\right)$ into consideration. Particularly, the maximum entropy objective function is defined as:
\begin{equation}
J(\pi)=\sum_{t=0}^{T} \mathbb{E}_{\left(\boldsymbol{s}_{t}, \boldsymbol{a}_{t}\right) \sim \rho_{\pi}}\left[r\left(\boldsymbol{s}_{t}, \boldsymbol{a}_{t}\right)+\alpha \mathcal{H}\left(\pi\left(\cdot \mid \boldsymbol{s}_{t}\right)\right)\right],
\end{equation}
where $\alpha$ is a factor to determine the importance of entropy relative to the reward; $\left(\pi\left(\cdot \mid \boldsymbol{s}_{t}\right)\right)$ represents the probability distribution of any action taken after the state $\mathbf{s}_{t}$, and the Gaussian probability distribution is generally used; $\mathcal{H}\left(\pi\left(\cdot \mid \boldsymbol{s}_{t}\right)\right)$ is the entropy term which is defined as $\mathcal{H}\left(\pi\left(\cdot \mid \boldsymbol{s}_{t}\right)\right)\triangleq \mathbb{E}_{\boldsymbol{a}}\left[-\log \left(\pi\left(\boldsymbol{a} \mid \boldsymbol{s}_{t}\right)\right)\right]$. It is worth noting that in the specific states, the agent will explore as many different actions as possible to maximize the target entropy. This strategy increases the exploratory nature of SAC.

Accordingly, the state value function $Q\left(\boldsymbol{s}_{t}, \boldsymbol{a}_{t}\right)$  and the action-state value function $V\left(\boldsymbol{s}_{t}\right)$ of the SAC can be defined as follows:
\begin{equation}
 Q\left(\boldsymbol{s}_{t}, \boldsymbol{a}_{t}\right) =r\left(\boldsymbol{s}_{t}, \boldsymbol{a}_{t}\right)+\gamma \mathbb{E}_{\boldsymbol{s}_{t+1} \sim p\left(\boldsymbol{s}_{t+1}, \tau \mid \boldsymbol{s}_{t}, \boldsymbol{a}_{t}\right)}\left[V\left(\boldsymbol{s}_{t+1}\right)\right],
\end{equation}
\begin{equation}
\quad\quad V\left(\boldsymbol{s}_{t}\right) =\mathbb{E}_{\boldsymbol{a}_{t} \sim \pi}\left[Q\left(\boldsymbol{s}_{t}, a_{t}\right)-\alpha \log \pi\left(\boldsymbol{a}_{t} \mid \boldsymbol{s}_{t}\right)\right].
\end{equation}

In the step of the policy improvement, the new policy is updated in the exponential direction of the new Q-function. The option of update can lead to policy improvements in term of the soft value. In the actual situations, we prefer tractable policies. Thus, we additionally limit the policy to a set of policies $\Pi$. Considering the constraint that $\pi \in \Pi$, the improved policy is projected into the desired policies set. For simplicity, we use the information projection defined by Kullback-Leibler divergence. Thus, we update the policy according to
\begin{equation}
\pi_{\text {new }}=\arg \min _{\pi^{\prime} \in \Pi} D_\mathrm{K L}\left(\pi^{\prime}\left(\cdot \mid \boldsymbol{s}_{t}\right) \bigg \| \frac{\exp \left(Q^{\pi_{\text {old }}}\left(\boldsymbol{s}_{t}, \cdot\right)\right)}{Z^{\pi_{\text {old }}\left(\boldsymbol{s}_{t}\right)}}\right),
\end{equation}
where $Z^{\pi_{\text {old }}\left(\boldsymbol{s}_{t}\right)}$  is adopted to normalize the distribution.

In order to solve the problem of the variables in continuous domain, we use a function approximator to represent the Q-values. Then, the SAC algorithm generates a target network of the policy $\pi$ and action-state value function $Q\left(\boldsymbol{s}_{t}, \boldsymbol{a}_{t}\right)$ for soft update, which can significantly improve the stability of learning. In the SAC framework, the agent can learn stochastic policies by maximizing the entropy objective functions which are expressed as value and policy functions. Therefore, in the value function, it encourages exploration by increasing the value of high-entropy actions. In the policy function, it can prevent the policy from converging early.
 Detailed update process of these functions will be introduced in the next sub-section.
\subsection{The Architecture and Update Process of SAC}
As discussed above, we utilize function approximators for the policy function, V-function and Q-function, and adopt stochastic gradient descent to alternately optimize the networks. We consider a tractable policy $\pi_{\phi}\left(\boldsymbol{a}_{t} \mid \boldsymbol{s}_{t}\right)$, a parameterized state value function $V_{\psi}\left(\boldsymbol{s}_{t}\right)$, and soft Q-function $Q_{\theta}\left(\boldsymbol{s}_{t}, \boldsymbol{a}_{t}\right)$. $\phi$, $\psi$, and $\theta$ are the
\begin{figure}[t]
\centering
\includegraphics[width = 3.33 in]{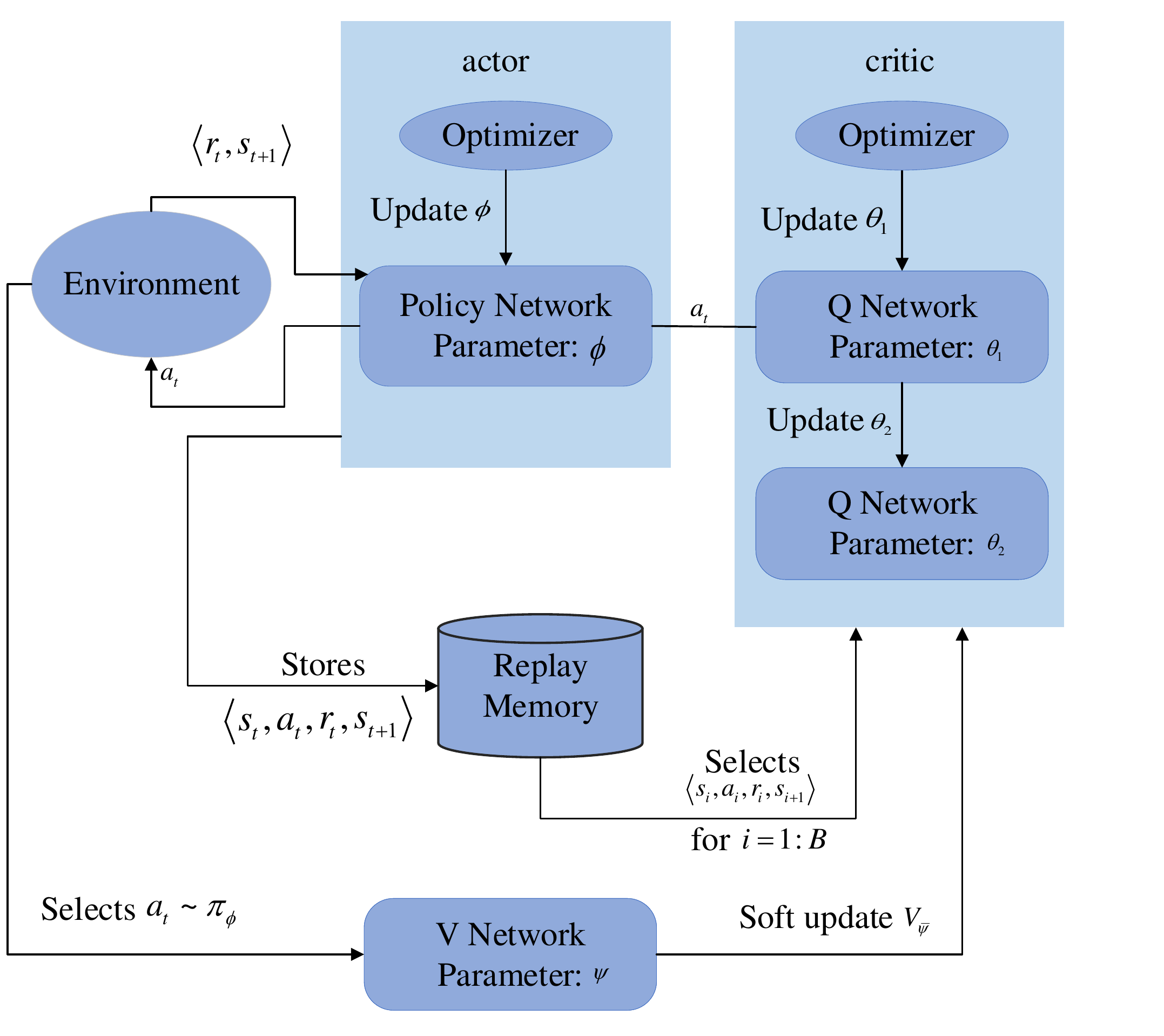}
\vspace{-0.1 cm}
\caption{The network structure of the Soft Actor-Critic.}
\vspace{-0.4 cm}
\end{figure}
parameters of these networks. For example, the policy can be modeled as a Gaussian distribution.
The complete architecture of the SAC framework is shown in Fig. 3. There are three types of DNN in our algorithm, namely V-network, policy network and Q-network. We also use two Q-networks to alleviate positive bias in the step of policy improvement. In particular, we use the parameters $\theta_{i}$ to parameterize two Q-functions, and train them separately to optimize $J_{Q}\left(\theta_{i}\right)$. Next, we will derive the update of these parameter vectors.

Firstly, the update of the soft value comes from the approximation of the state value function. The soft value function is trained by minimizing the squared residual error

\begin{equation}
\small
\begin{aligned}
&J_{V}(\psi)=\mathbb{E}_{\boldsymbol{s}_{t} \sim \mathcal{D}}\bigg [\frac{1}{2}\bigg ((V_{\psi}\left(\boldsymbol{s}_{t}\right)\\
&\quad \quad \quad-\mathbb{E}_{\boldsymbol{a}_{t} \sim \pi_{\phi}}\left[Q_{\theta}\left(\boldsymbol{s}_{t}, \boldsymbol{a}_{t}\right)
-\log \pi_{\phi}\left(\boldsymbol{a}_{t} \mid \boldsymbol{s}_{t}\right)\right]\bigg )^{2}\bigg ],\label{eq:v}
\end{aligned}
\end{equation}
where $\mathcal{D}$ is a replay buffer. Then, the gradient of the equation (\ref{eq:v}) is estimated using an unbiased estimator

\begin{equation}
\begin{aligned}
&\hat{\nabla}_{\psi} J_{V}(\psi)=\nabla_{\psi} V_{\psi}\left(\boldsymbol{s}_{t}\right)\text{(}V_{\psi}\left(\boldsymbol{s}_{t}\right)\\
&\quad\quad \quad\quad -Q_{\theta}\left(\boldsymbol{s}_{t}, \boldsymbol{a}_{t}\right)+\log \pi_{\phi}\left(\boldsymbol{a}_{t} \mid \boldsymbol{s}_{t}\right)\text { ) },\label{eq:1}
\end{aligned}
\end{equation}
where the action is selected from the current set of policies, instead of the replay buffer.

Secondly, the soft Q-function parameter is trained by minimizing the soft Bellman residual, which is defined as:
\begin{equation}
\small
J_{Q}(\theta)=\mathbb{E}_{\left(\boldsymbol{s}_{t}, \boldsymbol{a}_{t}\right) \sim \mathcal{D}}\left[\frac{1}{2}\left(Q_{\theta}\left(\boldsymbol{s}_{t}, \boldsymbol{a}_{t}\right)-\hat{Q}\left(\boldsymbol{s}_{t}, \boldsymbol{a}_{t}\right)\right)^{2}\right],\label{eq:Q}
\end{equation}
with $\hat{Q}\left(\boldsymbol{s}_{t}, \boldsymbol{a}_{t}\right)=r\left(\boldsymbol{s}_{t}, \boldsymbol{a}_{t}\right)+\gamma \mathbb{E}_{\boldsymbol{s}_{t+1} \sim p}\left[V_{\bar{\psi}}\left(\boldsymbol{s}_{t+1}\right)\right]$. The gradient of the equation (\ref{eq:Q}) is optimized with stochastic gradients
\begin{equation}
\begin{aligned}
&\hat{\nabla}_{\theta} J_{Q}(\theta)=\nabla_{\theta} Q_{\theta}\left(\boldsymbol{a}_{t}, \boldsymbol{s}_{t}\right)\text { ( }Q_{\theta}\left(\boldsymbol{s}_{t}, \boldsymbol{a}_{t}\right)\\
&\quad \quad \quad\quad-r\left(\boldsymbol{s}_{t}, \boldsymbol{a}_{t}\right)-\gamma V_{\bar{\psi}}\left(\boldsymbol{s}_{t+1}\right)\text { ) },\label{eq:2}
\end{aligned}
\end{equation}
where a target value network $V_{\bar{\psi}}$ is used for update. The parameter $\bar{\psi}$ is an exponentially moving average of the target value network weight, which is given by
\begin{equation}
\bar{\psi} \leftarrow \tau \psi+(1-\tau) \bar{\psi},
\end{equation}
where $\tau$ is a target smoothing coefficient to improve stability.

Finally, the policy parameter is learned by minimizing the expected KL-divergence:
\begin{equation}
\small
J_{\pi}(\phi)=\mathbb{E}_{\boldsymbol{s}_{t} \sim \mathcal{D}}\left[\mathrm{D}_{\mathrm{KL}}\left(\pi_{\phi}\left(\cdot \mid \boldsymbol{s}_{t}\right) \bigg \| \frac{\exp \left(Q_{\theta}\left(\boldsymbol{s}_{t}, \cdot\right)\right)}{Z_{\theta}\left(\boldsymbol{s}_{t}\right)}\right)\right].
\end{equation}
For simplicity, we adopt neural network transformation to reparameterize the policy
\begin{equation}
\boldsymbol{a}_{t}=f_{\phi}\left(\epsilon_{t} ; \boldsymbol{s}_{t}\right),
\end{equation}
where $\epsilon_{t} $ is a noise vector. The objective can be rewritten as
\begin{equation}
\begin{aligned}
&J_{\pi}(\phi)=\mathbb{E}_{\boldsymbol{s}_{t} \sim \mathcal{D}, \epsilon_{t} \sim \mathcal{N}}\big[\log \pi_{\phi}\left(f_{\phi}\left(\epsilon_{t} ; \boldsymbol{s}_{t}\right) \mid \boldsymbol{s}_{t}\right)\\
&\quad \quad \quad-Q_{\theta}\left(\boldsymbol{s}_{t}, f_{\phi}\left(\epsilon_{t} ; \boldsymbol{s}_{t}\right)\right)\big].\label{eq:PI}
\end{aligned}
\end{equation}
Next, the gradient of the above equation (\ref{eq:PI}) can be approximated as
\begin{equation}
\begin{aligned}
&\hat{\nabla}_{\phi} J_{\pi}(\phi)=\nabla_{\phi} \log \pi_{\phi}\left(\boldsymbol{a}_{t} \mid \boldsymbol{s}_{t}\right) \\
&\quad+\left(\nabla_{\boldsymbol{a}_{t}} \log \pi_{\phi}\left(\boldsymbol{a}_{t} \mid \boldsymbol{s}_{t}\right)-\nabla_{\boldsymbol{a}_{t}} Q\left(\boldsymbol{s}_{t}, \boldsymbol{a}_{t}\right)\right) \nabla_{\phi} f_{\phi}\left(\epsilon_{t} ; \boldsymbol{s}_{t}\right).\label{eq:3}
\end{aligned}
\end{equation}
The unbiased gradient estimator extends the deterministic policy gradients to stochastic policies.

\subsection{Digital Beamformer Design}

In each episode, we can find the optimal $\boldsymbol{a}_{t}$, and calculate the optimal $\mathbf{F}_{\mathrm{RF}}^{*}$ and $\mathbf{\Phi}^{*}$. Then, the effective channel is assumed as
\begin{equation}
\mathbf{H}_\mathrm{e f f}=\left(\mathbf{H}^{H} \mathbf{\Phi} \mathbf{H}_\mathbf{r}\right) \mathbf{F}_\mathrm{R F},\label{eq:heff}
\end{equation}
where $\mathbf{H}_\mathbf{r}=\left[\mathbf{h}_{1}, \ldots,  \mathbf{h}_{K}\right]$. Thus, we design the digital beamformer via MMSE method as:
\begin{equation}
\mathbf{F}_\mathrm{B B}^{*}=\left((\mathbf{H}_\mathrm{e f f}) \mathbf{H}_\mathrm{e f f}^{H}+\left(\frac{\sigma^{2} }{P}
 \right) \left(\mathbf{F}_\mathrm{R F}\right) \mathbf{F}_\mathrm{R F}^{H}\right)^{-1} \mathbf{H}_\mathrm{e f f}.\label{eq:fbb}
\end{equation}
Finally, to guarantee the power constraint, the final digital beamformer is normalized
\begin{equation}
\mathbf{F}_\mathrm{B B}^{*}=\frac{\sqrt{N_{s}} \mathbf{F}_\mathrm{B B}^{*}}{\left\|\mathbf{F}_\mathrm{R F}^{*} \mathbf{F}_\mathrm{B B}^{*}\right\|_{F}}.\label{eq:FBB}
\end{equation}

Thus, the above SAC based jointly design of hybrid beamforming and passive beamforming framework is summarized as Algorithm I.
\begin{algorithm}[t]
\small
\caption{SAC-based Active Hybrid Beamforming and Passive Beamforming Design}
\label{alg:1}
\begin{algorithmic}[1]
 \REQUIRE   $\theta_{1}$, $\theta_{2}$, $\psi$, $\phi$

 \textbf{Analog precoder and RIS PSs design}
\STATE{Initialize parameter vectors $\theta_{1}$, $\theta_{2}$, $\psi$, $\phi$}
\STATE{Initialize experience memory $\mathcal{D}$}

            \FOR {each episode }
                \STATE{Initialize state $\boldsymbol{s}_{0} \in \boldsymbol{S}, \boldsymbol{s} \leftarrow \boldsymbol{s}_{0}$}
                \FOR {each step }
                \STATE{$\boldsymbol{a}_{t} \sim \pi_{\phi}\left(\boldsymbol{a}_{t} \mid \boldsymbol{s}_{t}\right)$}
                \STATE{$\boldsymbol{s}_{t+1} \sim p\left(\boldsymbol{s}_{t+1} \mid \boldsymbol{s}_{t}, \boldsymbol{a}_{t}\right)$}
                \STATE{$\mathcal{D} \leftarrow \mathcal{D} \cup\left\{\left(\boldsymbol{s}_{t}, \boldsymbol{a}_{t}, \boldsymbol{r}_{t}, \boldsymbol{s}_{t+1}\right)\right\}$}
                \STATE{Sample from $\mathcal{D}$ and compute $\nabla J_{Q}(\theta_{i}), i \in\{1,2\}$ by using (\ref{eq:2})}
                \STATE{Update Q-networks parameters, $\theta_{i} \leftarrow \theta_{i}-\lambda_{Q} \hat{\nabla}_{\theta_{i}} J_{Q}\left(\theta_{i}\right)$ for $i \in\{1,2\}$}
                \STATE{Sample from the fixed distribution and compute $\nabla J_{\pi}(\phi)$  by using (\ref{eq:3})}
                \STATE{Update policy network parameter, $\phi \leftarrow \phi-\lambda_{\pi} \hat{\nabla}_{\phi} J_{\pi}(\phi)$ }
                \STATE{Sample from current policy and compute $\hat{\nabla}_{\psi} J_{V}(\psi)$ by using (\ref{eq:1})}
                \STATE{Update V network parameter, $\psi \leftarrow \psi-\lambda_{V} \hat{\nabla}_{\psi} J_{V}(\psi)$ }
                \STATE{$\bar{\psi} \leftarrow \tau \psi+(1-\tau) \bar{\psi}$}
                \STATE{Update the next state $\boldsymbol{s}_{t} \leftarrow \boldsymbol{s}_{t+1}$}

                  \textbf{Digital precoder design}
                 \STATE{Select the optimal action to get $\mathbf{F}_{\mathrm{RF}}$ and $\mathbf{\Phi}$ }
                  \STATE{Obtain the effective channel $\mathbf{H}_\mathrm{e f f}$ by using (\ref{eq:heff}) }
                   \STATE{Compute and normalize the optimal digital precoder  $\mathbf{F}_{\mathrm{BB}}^{*}$ by using (\ref{eq:fbb}) and (\ref{eq:FBB})}
                \ENDFOR
               \ENDFOR
                \ENSURE  $\mathbf{F}_{\mathrm{RF}}^{*}$,   $\mathbf{\Phi}^{*}$, $\mathbf{F}_{\mathrm{BB}}^{*}$
   \end{algorithmic}
\end{algorithm}

\section{Simulation Results}
\vspace{-0 cm}
In this section, we present numerical results of our proposed SAC based joint hybrid beamforming and passive beamforming design for the RIS assisted MU-MISO system. We assume the BS has $N_\mathrm{t}=32$ antennas and $N_\mathrm{RF}=3$ RF chains to serve $K=3$ users. In the mmWave channel model, the number of propagation paths $L$ is set as 4. We define the signal-to-noise-ratio as $\mathrm{SNR} = \frac{P}{{\sigma}^2}$, where ${\sigma}^2=1$.  In addition, the hyperparameters of the proposed SAC scheme is summarized in Table I.
For comparison purposes, we also evaluate the deterministic policy based SAC (DP based SAC), in which the entropy target $\alpha$ = 0. It means that the influence of the stochastic policy brought to the policy update is not considered.  Besides, we adopt the state-of-the-art DDPG algorithm for comparison.


    \begin{table}[t]
   \small
		\centering
        \caption{SAC hyperparameters}
		\begin{tabular}{l l}
        \hline
        Hyperparameter&Value\\\hline
		Layers&	2 fully connected layers\\
		Layer hidden units	&  256\\
        Activation function  &  ReLU\\
        Batch size  &  64\\
        Replay buffer size  &  1000000\\
        Target smoothing coefficient  &   0.005\\
        Target update interval & 1\\
        Discount rate  & 0.95 \\
        Learning iterations per round  & 1 \\
        Learning rate  & 0.0001 \\
        Optimizer  & Adam  \\
        Loss  & Mean squared error  \\
        Entropy target factor $\alpha$  & 0.2 \\
        \hline
		\end{tabular}
		\label{tab:Margin_settings}
	\end{table}

\begin{figure}
\centering
\includegraphics[width = 2.7 in]{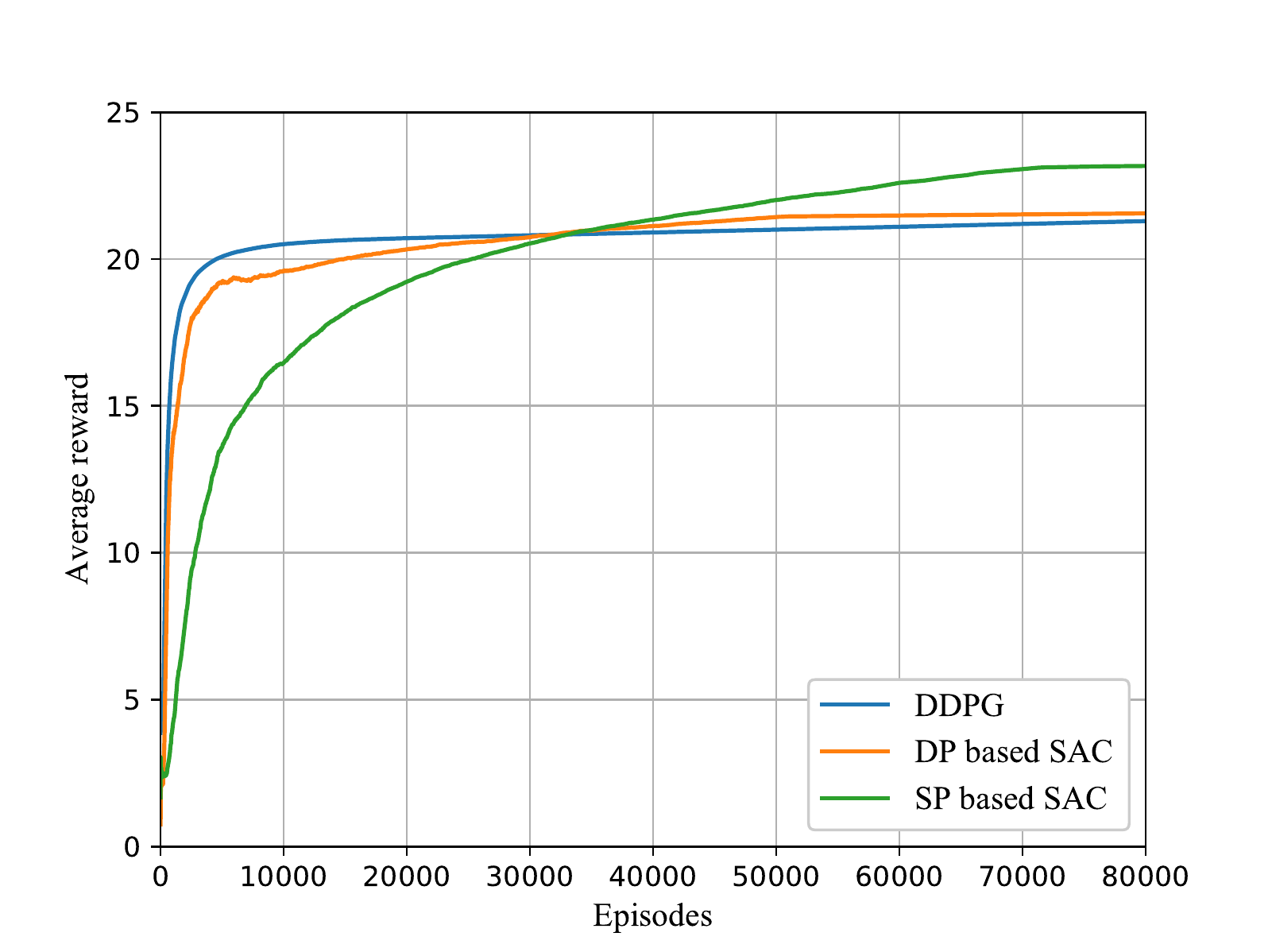}
\vspace{-0.1 cm}
\caption{Average reward versus episodes ($N_\mathrm{t}=32, M = 64$, SNR = 10dB). }
\label{fig:bb}
\vspace{-0.6 cm}
\end{figure}

\begin{figure}[t]
\centering
\includegraphics[width = 2.7 in]{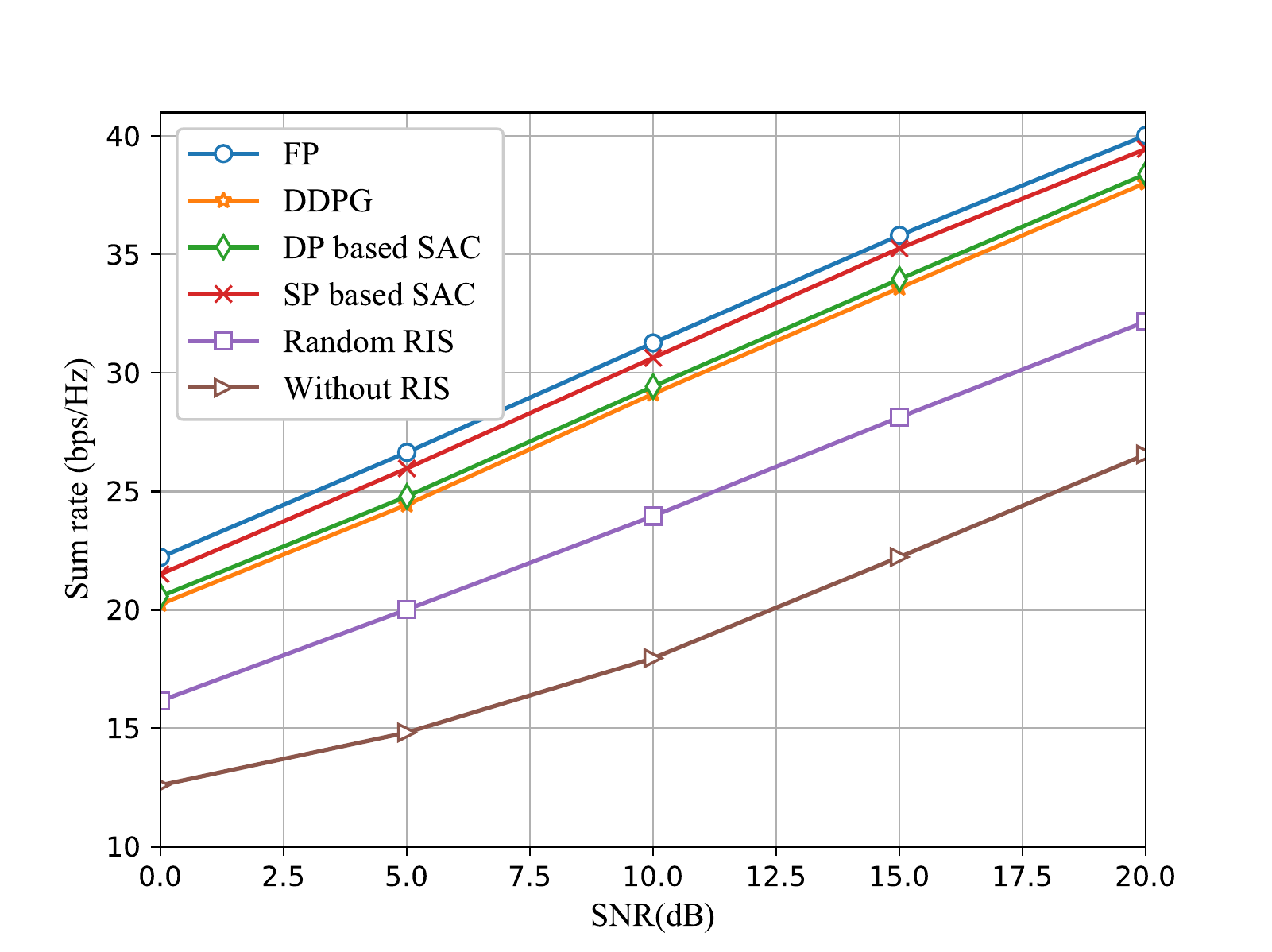}
\vspace{-0.1 cm}
\caption{Sum-rate versus SNR for different schemes ($N_\mathrm{t}=32, M=64$ ).}
\label{fig:cc}
\vspace{0.1 cm}
\end{figure}
In order to demonstrate the learning process, we present the average reward versus learning episodes in  Fig. \ref{fig:bb}, where $M$ = 64 and SNR is set as 10dB. We can see that the proposed stochastic policy based SAC approach requires more episodes to convergence compared to other deterministic policy based approaches. This is because the stochastic policy based scheme enables the agent to explore more stochastic actions in a certain state. When all the approaches converge, the proposed stochastic policy based SAC has the better performance. Besides, DP based SAC has a similar trend as DDPG. Since the agent can choose the deterministic strategy in the early learning stage, they converge quickly to a poor local optimum.

In order to better evaluate the performance of our proposed algorithm, we consider three additional  benchmark schemes: 1) FP: an iterative algorithm based on fractional programming \cite{21} to design fully digital beamformer and phase-shift of RIS; 2) Random RIS: $\mathbf{\Phi}$ is designed randomly; 3) Without RIS: RIS is not deployed in the system.
Fig. \ref{fig:cc} shows the sum-rate versus SNR over the different schemes, where the RIS has $M=64 $ elements.
 Our proposed SAC algorithm based stochastic strategy outperforms the other two deterministic strategy based algorithms. It can be seen that our proposed SAC-based algorithm obtains
the comparable sum-rate performance with fractional programming algorithm.  Besides, we can see that the system with the RIS achieves significantly a higher sum-rate than the system with randomly RIS and without RIS, which demonstrates the advantage of RIS in mmWave communication systems.

\begin{figure}[t]
\centering
\includegraphics[width = 2.61 in]{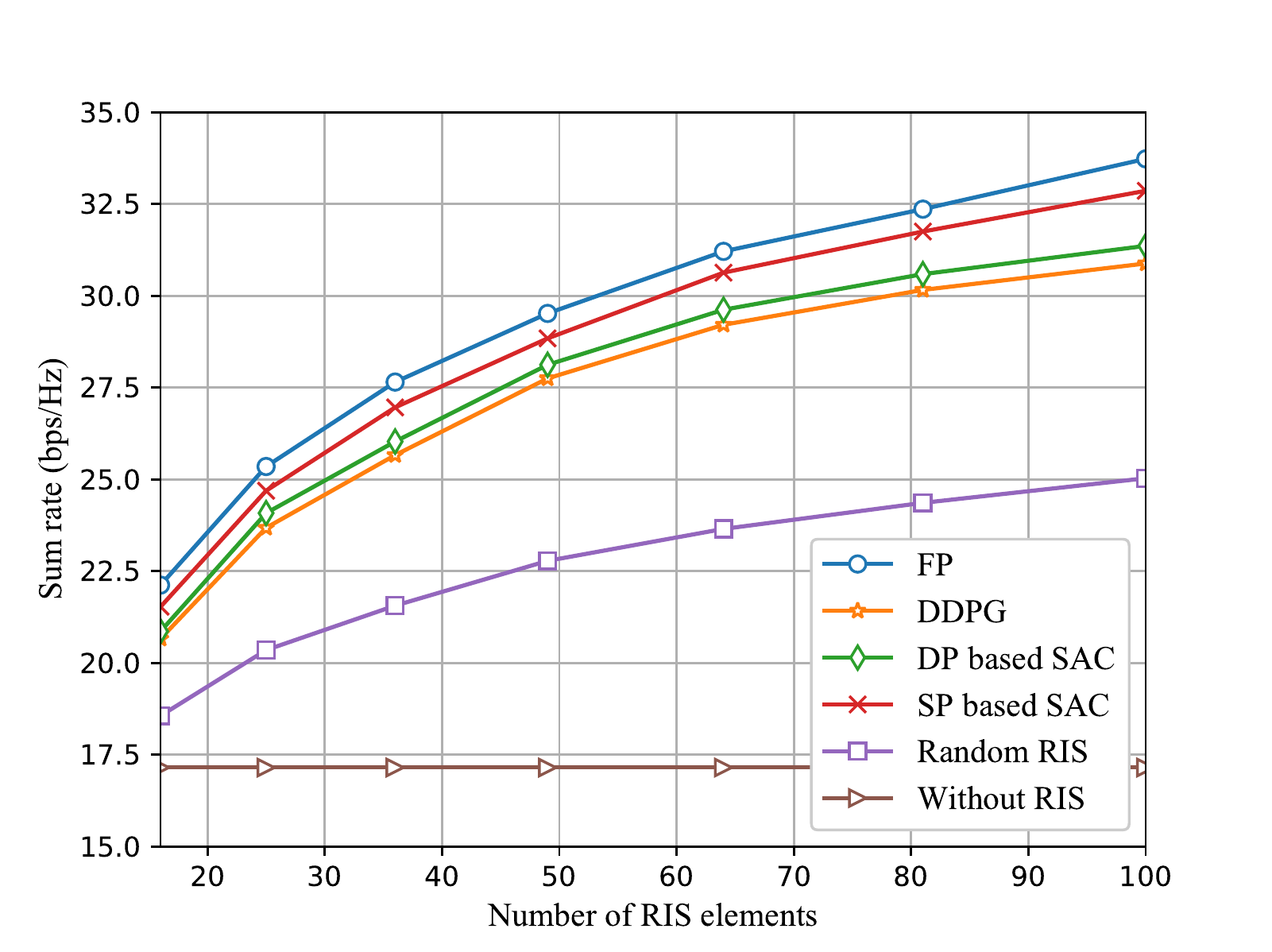}
\vspace{0.6 cm}
\caption{Sum-rate versus the number of RIS elements ($N_\mathrm{t}=32$, SNR = 10dB).}
\label{fig:aa}
\vspace{-0.5 cm}
\end{figure}


Finally, Fig. \ref{fig:aa} shows the sum-rate versus the number of RIS elements, where SNR = 10dB. We see that the sum-rate increases with the growing of the number of RIS elements. In addition, our proposed SAC-based algorithm obtains the sum-rate performance close to the full-digital FP algorithm with the growing of the number of RIS elements.
 On the other hand, our proposed SAC scheme outperforms DDPG and other benchmarks. As the number of RIS elements increases, the interval between the two algorithms becomes larger. This is because that as the dimension of the variable increases, the randomly selected strategy can explore more actions and find better policies compared to the deterministic policy.

\section{Conclusions}
\vspace{0.2 cm}

In this paper, we considered a RIS-assisted multi-user multiple-input single-output (MU-MISO) mmWave system and attempted to utilize a deep reinforcement learning (DRL) framework to jointly design active hybrid beamformer and passive beamformer.
We employed a soft actor-critic (SAC) algorithm in the DRL framework to jointly design active analog precoder and passive beamformer. Different from the traditional DRL algorithms, the SAC algorithm can explore more effective and better combination strategies through continuous random selection of strategies. After obtaining the active analog precoder and passive beamformer, the digital precoder is designed by minimum mean square error (MMSE) method.
The experimental results demonstrated that our proposed SAC-based DRL algorithm can achieve better performance compared with conventional DDPG algorithm.


\enlargethispage{-6.5cm}

\begin{thebibliography}{t}
\bibitem{7446253} S. Buzzi, C.-L. I, T. E. Klein, H. V. Poor, C. Yang, and A. Zappone, ``A survey of energy-efficient techniques for 5G networks and challenges ahead,"
\textit{IEEE J. Sel. Areas Commun.}, vol. 34, no. 4, pp. 697-709, Apr. 2016.


\bibitem{9326394} Q. Wu, S. Zhang, B. Zheng, C. You, and R. Zhang, ``Intelligent reflecting surface-aided wireless communications: A tutorial," \textit{IEEE Trans. Wireless Commun.}, vol. 69, no. 5, pp. 3313-3351, May 2021.

\bibitem{8910627}Q. Wu and R. Zhang, ``Towards smart and reconfigurable environment: Intelligent reflecting surface aided wireless network," \textit{IEEE Commun. Mag.}, vol. 58, no. 1, pp. 106-112, Nov. 2020.

\bibitem{8855810}X. Yu, D. Xu, and R. Schober, ``MISO wireless communication systems via intelligent reflecting surfaces,'' in \textit{Proc. IEEE Int. Conf. Commun. China (ICCC)}, Changchun, China, Aug. 2019, pp 735-740.

\bibitem{9148947} J. Wang, Y.-C. Liang, S. Han, and Y. Pei, ``Robust beamforming and
phase shift design for IRS-enhanced multi-user MISO downlink communication," in \textit{Proc. IEEE Int. Conf. Commun. (ICC)}, Dublin, Ireland, Jun. 2020, pp. 1-6.


\bibitem{21} H. Guo, Y.-C. Liang, J. Chen and E. G. Larsson, ``Weighted Sum-Rate Maximization for Intelligent Reflecting Surface Enhanced Wireless Networks,'' in \textit{Proc. IEEE Global Commun. Conf. (GLOBECOM)}, Waikoloa Village, HI, Dec. 2019, pp. 1-6.

\bibitem{xiu2021sumrate}Y. Xiu, W. Sun, J. Wu, G. Gui, N. Wei, and Z. Zhang, ``Sum-rate maximization in distributed intelligent reflecting surfaces-aided mmwave communications," Jan. 2021. [Online]. Available: https://arxiv.org/abs/2101.07073.

\bibitem{8938771}F. B. Mismar, B. L. Evans, and A. Alkhateeb, ``Deep reinforcement learning for 5G networks: Joint beamforming, power control, and interference coordination," \textit{IEEE Trans. Commun.}, vol. 68, no. 3, pp. 1581-1592, Mar. 2020.

\bibitem{9110869}C. Huang, R. Mo, and C. Yuen, ``Reconfigurable intelligent surface assisted multiuser MISO systems exploiting deep reinforcement learning," \textit{IEEE J. Sel. Areas Commun.}, vol. 38, no. 8, pp. 1839-1850, Aug. 2020.

\bibitem{zhang2020millimeter} Q. Zhang, W. Saad and M. Bennis, ``Millimeter Wave communications with an intelligent reflector: Performance optimization and distributional reinforcement learning,'' \textit{IEEE Trans. Wireless Commun.}, to appear.

\bibitem{8968350}K. Feng, Q. Wang, X. Li, and C.-K. Wen, ``Deep reinforcement learning based intelligent reflecting surface optimization for MISO communication systems," \textit{IEEE Commun. Lett.}, vol. 9, no. 5, pp. 745-749, Jan. 2020.

\bibitem{9322372}J. Lin, Y. Zout, X. Dong, S. Gong, D. T. Hoang, and D. Niyato, ``Deep reinforcement learning for robust beamforming in IRS-assisted wireless communications," in \textit{Proc. IEEE Global Commun. (GLOBECOM)}, Taipei, Taiwan, Dec. 2020, pp. 1-6.

\bibitem{channel}M. Xiao, S. Mumtaz, Y. Huang, L. Dai, Y. Li, M. Matthaiou, G. K. Karagiannidis, E. Bj\"{o}rnson, K. Yang, C.-L. I, and A. Ghosh, ``Millimeter wave communications for future mobile networks," \textit{IEEE J. Sel. Areas Commun.}, vol. 35, no. 9, pp. 1909-1935, Sep. 2017.

\bibitem{haarnoja2018soft}T. Haarnoja, A. Zhou, P. Abbeel, and S. Levine,``Soft actor-critic: Off-policy maximum entropy deep reinforcement learning with a stochastic actor," in \textit{Proc. IEEE  Int. Conf. Machine Learning (ICML)}, Stockholm Sweden, Jul. 2018, pp. 2976-2989.




\end{thebibliography}
\end{document}